\documentclass[twocolumn]{pasj01}

\begin{document}
\SetRunningHead{Hitomi Collaboration}{Hitomi X-ray Observation of the Pulsar Wind Nebula G21.5$-$0.9}
\Received{2017/12/13}
\Accepted{2018/2/14}

\title{Hitomi X-ray Observation of the Pulsar Wind Nebula G21.5$-$0.9
\thanks{The corresponding authors are
Hiroyuki \textsc{Uchida},
Takaaki \textsc{Tanaka},
Samar \textsc{Safi-Harb},
Masahiro \textsc{Tsujimoto},
Yukikatsu \textsc{Terada},
Aya \textsc{Bamba},
Yoshitomo \textsc{Maeda},
and
John P. \textsc{Hughes}
}}


\author{Hitomi Collaboration,
Felix \textsc{Aharonian}\altaffilmark{1,2,3},
Hiroki \textsc{Akamatsu}\altaffilmark{4},
Fumie \textsc{Akimoto}\altaffilmark{5},
Steven W. \textsc{Allen}\altaffilmark{6,7,8},
Lorella \textsc{Angelini}\altaffilmark{9},
Marc \textsc{Audard}\altaffilmark{10},
Hisamitsu \textsc{Awaki}\altaffilmark{11},
Magnus \textsc{Axelsson}\altaffilmark{12},
Aya \textsc{Bamba}\altaffilmark{13,14},
Marshall W. \textsc{Bautz}\altaffilmark{15}
% ...
%
}

\altaffiltext{1}{Dublin Institute for Advanced Studies, 31 Fitzwilliam Place, Dublin 2, Ireland}
\altaffiltext{2}{Max-Planck-Institut f{\"u}r Kernphysik, P.O. Box 103980, 69029 Heidelberg, Germany}
\altaffiltext{3}{Gran Sasso Science Institute, viale Francesco Crispi, 7 67100 L'Aquila (AQ), Italy}
\altaffiltext{4}{SRON Netherlands Institute for Space Research, Sorbonnelaan 2, 3584 CA Utrecht, The Netherlands}
\altaffiltext{5}{Institute for Space-Earth Environmental Research, Nagoya University, Furo-cho, Chikusa-ku, Nagoya, Aichi 464-8601}
\altaffiltext{6}{Kavli Institute for Particle Astrophysics and Cosmology, Stanford University, 452 Lomita Mall, Stanford, CA 94305, USA}
\altaffiltext{7}{Department of Physics, Stanford University, 382 Via Pueblo Mall, Stanford, CA 94305, USA}
\altaffiltext{8}{SLAC National Accelerator Laboratory, 2575 Sand Hill Road, Menlo Park, CA 94025, USA}
\altaffiltext{9}{NASA, Goddard Space Flight Center, 8800 Greenbelt Road, Greenbelt, MD 20771, USA}
\altaffiltext{10}{Department of Astronomy, University of Geneva, ch. d'\'Ecogia 16, CH-1290 Versoix, Switzerland}
\altaffiltext{11}{Department of Physics, Ehime University, Bunkyo-cho, Matsuyama, Ehime 790-8577}
\altaffiltext{12}{Department of Physics and Oskar Klein Center, Stockholm University, 106 91 Stockholm, Sweden}
\altaffiltext{13}{Department of Physics, The University of Tokyo, 7-3-1 Hongo, Bunkyo-ku, Tokyo 113-0033}
\altaffiltext{14}{Research Center for the Early Universe, School of Science, The University of Tokyo, 7-3-1 Hongo, Bunkyo-ku, Tokyo 113-0033}
\altaffiltext{15}{Kavli Institute for Astrophysics and Space Research, Massachusetts Institute of Technology, 77 Massachusetts Avenue, Cambridge, MA 02139, USA}
% ...

\email{uchida@cr.scphys.kyoto-u.ac.jp}


\maketitle

\begin{abstract}
We present results from the Hitomi X-ray observation of a young composite-type supernova remnant (SNR) G21.5$-$0.9, whose emission is dominated by the pulsar wind nebula (PWN) contribution. 
The X-ray spectra in the 0.8--80~keV range obtained with the Soft X-ray Spectrometer (SXS), Soft X-ray Imager (SXI) and Hard X-ray Imager (HXI)  show a significant break in the continuum as previously found with the NuSTAR observation. 
After taking into account all known emissions from the SNR other than the PWN itself, we find that the Hitomi spectra can be fitted with a broken power law with photon indices of $\Gamma_1=1.74\pm0.02$ and $\Gamma_2=2.14\pm0.01$ below and above the break at $7.1\pm0.3$~keV, which is significantly lower than the NuSTAR result ($\sim9.0$~keV).
The spectral break cannot be reproduced by time-dependent particle injection one-zone spectral energy distribution models, which strongly indicates that a more complex emission model is needed, as suggested by  recent theoretical models.
We also search for narrow emission or absorption lines with the SXS, and perform a timing analysis of PSR~J1833$-$1034 with the  HXI and SGD. 
No significant pulsation is found from the pulsar. 
However, unexpectedly, narrow absorption line features are detected in the SXS data at 4.2345~keV and 9.296~keV with a significance of 3.65~$\sigma$. 
While the origin of these features is not understood, their mere detection opens up a new field of research and was only possible with the high resolution, sensitivity and ability to measure 
extended sources provided by an X-ray microcalorimeter. 
\end{abstract}

\section{Introduction}
A pulsar wind nebula (PWN) is driven by relativistic particles and magnetic field generated by its central compact object, a pulsar inside a supernova remnant (SNR) shell \citep{Pacini1973, Rees1974, Kennel1984}.
A bubble is formed beyond a termination shock where the relativistic wind of non-thermal electrons and positrons interact with the surrounding ejecta (e.g., \cite{Fang2010}).
The resultant emission is dominated by centrally peaked synchrotron radiation from radio to X-rays and inverse Compton scattering (IC) at higher energies.
The observed spectra of PWNe are basically characterized by a power law with a hard spectral index $\alpha\sim-0.3-0$ at radio wavelengths and a steeper photon index in X-rays, $\Gamma\equiv1-\alpha\sim2$ (cf. \cite{Gaensler2006}).
Because the break energy is associated with the acceleration process and the aging of the particles, a wide-band analysis helps us understand the evolution of PWNe \citep{Reynolds1984}, although the nature of the spectral steepening is still under debate.

One of the best observed examples of a young PWN is G21.5$-0.9$ \citep{Altenhoff1970, Becker1981}, which substitutes for the Crab nebula \citep{Kirsch2005} as a standard candle or a calibration target for X-ray satellites.
Several X-ray studies of this nebula with Chandra and XMM-Newton show a non-thermal power-law spectrum with no line emission \citep{Slane2000, Safi-Harb2001, Warwick2001}.
Using G21.5$-$0.9, \citet{Tsujimoto2011} performed a comprehensive cross calibration of Chandra, INTEGRAL, RXTE, Suzaku, Swift, and XMM-Newton as one of the  activities of the International Astronomical Consortium for High Energy Calibration (IACHEC).
They separated these instruments into two groups; Chandra ACIS, Suzaku XIS, Swift XRT, and XMM-Newton EPIC (MOS and pn) for the soft band ($<10$~keV);  INTEGRAL IBIS-ISGRI, RXTE PCA, and Suzaku HXD-PIN for the hard band ($>10$~keV).
One of their results of interest to scientific studies is a significant difference of photon indices $\Gamma\sim1.84$ and $\sim2.05$ taken from the joint fittings of the soft- and hard-band instruments, respectively.
This study implies spectral steepening of G21.5$-$0.9 in the X-ray band, as indicated by the preceding soft-band analyses (e.g., \cite{Matheson2010}, in addition to the above), although the radially dependent $\Gamma$ should be considered in the discussion of the nature of the steepening.
\citet{Nynka2014} observed G21.5$-$0.9 with NuSTAR and revealed a high-energy spectral feature in the band of 3--45~keV.
The spectrum is represented by a broken power law with a break energy of $\sim9$~keV.
A broadband spectral energy distribution (SED) model built by \citet{Tanaka2011} gives a poor fit to the NuSTAR spectrum and thus \citet{Nynka2014} 
suggested that further modeling is required to explain the wide-band spectrum of G21.5$-$0.9.
They proposed some extra aspects to take into account, for example, more complex electron injection spectra, additional loss processes (e.g., diffusion) or radial dependence of the PWN parameters.

One of the clear differences between G21.5$-$0.9 and the Crab is
the existence of faint thin-thermal extended emission
\citep{Bocchino2005,Matheson2005,Matheson2010}.
This fact illustrates how accumulated calibration observations help to reveal a shell component in a Crab-like PWN.  
However, given the brightness of the PWN and the relatively weak thermal X-ray emission from G21.5$-$0.9,
the parameters of the thermal emission from the shell are still poorly determined.
In particular, we have no information on Fe-K emission line
which is common in young SNRs such as Cassiopeia~A \citep{hughes2000}. 
Depending on the magnetic field strength of the powering pulsar, the emission from the pulsar itself also reveals line features
in the X-ray band due to the cyclotron effect \citep{Meszaros1985}.
It is thus of interest to search for emission/absorption line structures
with excellent energy resolution detectors.

PSR~J1833$-$1034 was discovered at the center of G21.5$-$0.9 in the radio band \citep{gupta2005,Camilo2006} and GeV gamma-ray band \citep{abdo2013}.
The characteristic age of the pulsar is estimated to be 4850~yr from the period of $\sim61.9$~ms and the period derivative of $\sim2.0\times 10^{-13}$~s~s$^{-1}$, however the dynamics of its associated PWN indicates a much younger age of 870$^{+200}_{-150}$~yr \citep{Bietenholz2008}, which makes this pulsar one of the youngest and the most energetic systems in our Galaxy.
On the other hand, no significant pulsation has been found yet in the X-ray band \citep{Camilo2006,Bocchino2005,Matheson2010}, although the central pulsar is very energetic \citep{kargaltsev2008,bamba2010}.
It is likely due to the contamination from the very bright PWN.
Typically, X-ray emission from a pulsar is harder than that from the PWN \citep{kargaltsev2008}, and therefore, the hard X-ray band is suitable to search for the coherent pulsation.
Hitomi HXI has good sensitivity, low background \citep{HitomiHXI2017,HitomiHXT2017, HitomiHXIOrbit2017}, and good timing accuracy \citep{HitomiTimeSystem2017} with a rather long time duration of the G21.5$-$0.9 observation of 329~ks, and thus it could have higher sensitivity for the search for the coherent pulsation from the pulsar.

In this paper we report on observational results of G21.5$-$0.9 with Hitomi (formerly known as ASTRO-H; \cite{Takahashi2014}).
The observation was performed during the commissioning and performance verification phase.
We obtained simultaneous data of all the instruments aboard with the longest exposure among the targeted celestial sources Hitomi observed.
Here we focus on the following three studies; a wide-band spectroscopy, narrow emission or absorption line searches, and a timing analysis.
In section~2, we present detailed information on the Hitomi observation and the data reduction. 
In section~3, we perform the joint fitting of the G21.5$-$0.9 data and discuss the result.
The blind search of emission or absorption lines and the timing analysis are presented in sections~4 and 5, respectively.
All the results are summarized in section~6.

\begin{table*}[t]
  \caption{Observation log.}\label{tab:log}
  \begin{center}
    \begin{tabular}{lccccc}
\hline
Target & Obs. Date &  (R.A., Dec.)$_{\rm{J2000}}$ &  Sequence ID & Effective Exposure (ks)\\
\hline	   						    		      			  						
G21.5$-$0.9 & 2016 Mar 19--23 &  (278.39, $-$10.57)  & 100050010--100050040  & 165 (SXS) / 51 (SXI) / 99 (HXI) / 255 (SGD)\\
\hline
    \end{tabular}
  \end{center}
\end{table*}

\section{Observation and Data Reduction}
G21.5$-$0.9 was observed with Hitomi on 2016 March 19--23 during the instrument commissioning phase of the satellite.
We analyzed data from the four instruments aboard Hitomi: the Soft X-ray Spectrometer (SXS; \cite{Kelley2016}), the Soft X-ray Imager (SXI; \cite{Tanaka2017}), the Hard X-ray Imager (HXI; \cite{HitomiHXI2017}), and the Soft Gamma-ray Detector (SGD; \cite{Watanabe2016}). 
The Soft X-ray Telescope  (SXT; \cite{Soong2014, Okajima2016}) consists of two modules of X-ray mirrors, SXT-S and SXT-I, which focus X-rays for the SXS and SXI, respectively. 
The HXI system consists of two sets of detector modules referred to as HXI1 and HXI2. 
Two sets of the Hard X-ray Telescope (HXT; \cite{Awaki2014}) are used to focus hard-band X-rays for each of the HXI sensors. 
The SGD system consists of two sets of detector modules referred to as SGD1 and SGD2. 
Detailed information on the observation is summarized in table~\ref{tab:log}.

We combined all the data of four different sequence IDs (see table~\ref{tab:log}) for our spectral analysis.
We performed the data reduction with version 6.20 of the HEAsoft tools, which is compatible with version 005b of the Hitomi Software released on 2017 March 6.
We  applied the Hitomi Calibration Database version 6 released on 2017 March 6 for the following analysis. 
Note that the gate valve of the SXS remained closed during the observation, which significantly reduced the effective area of the SXS below 2~keV. 
We applied the ``Crab ratio correction factor'' for modeling the effective area of SXS \citep{Tsujimoto2017}.
In the SXI data analysis, we carefully excluded events detected in ``minus-Z day earth (MZDYE)'' intervals, during which the SXI has many pixels affected by light leakage from the day earth \citep{Nakajima2017}.
We eliminated the SGD data for the wide-band spectroscopy since the observation was performed during  the turn-on phase of SGD1 and we have no SGD2 data.

\begin{figure}[t]
 \begin{center}
  \FigureFile(8cm,){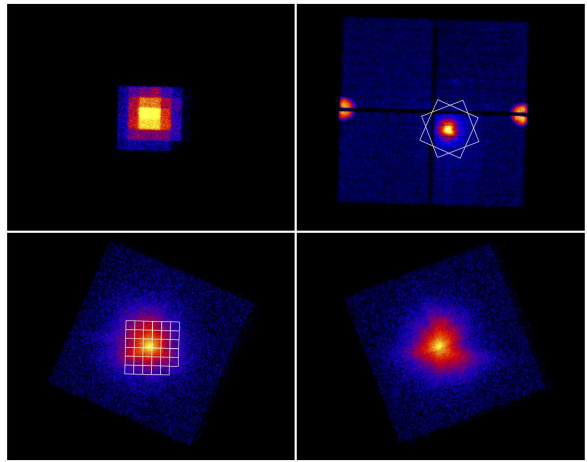} 
 \end{center}
\caption{\textit{Top left}: SXS sky coordinate image of G21.5$-$0.9. \textit{Top right}: SXI image of the source and the surrounding region. Two calibration-source regions are also included in the image.  The FOV of the HXI are indicated by the solid squares. \textit{Bottom}: HXI1 (left) and HXI2 (right) images of G21.5$-$0.9. The SXS pixel array (white squares) overlaid on the HXI1 image. }\label{fig:image}
\end{figure}

In figure~\ref{fig:image}, we present the full-band images of G21.5$-$0.9 taken by the SXS, SXI, and HXI.
We note that there are no significant transient sources in the vicinity of G21.5$-$0.9 within the field of view (FOV) of the SXI.
As previously reported by \citet{Slane2000}, G21.5$-$0.9 has a core of the wind termination shock surrounded by a synchrotron nebula with a radius of $\sim30\arcsec$, which is consistent with the centrally-peaked profile shown in figure~\ref{fig:image}.
G21.5$-$0.9 also has a faint $150\arcsec$  radius halo that almost covers the $3\arcmin\times3\arcmin$ SXS FOV.

To extract the SXS spectrum, we used all 35 pixels.
The source extraction region for the SXI and HXI is a circle with a $\sim3\arcmin$ radius  centered at $\rm{(R.A.,~Dec.)=(18^{h}~33^{m}~33\fs57,~-10\arcdeg~34\arcmin~07\farcs5)}$ in the equinox J2000.0, which is the position of the central pulsar, PSR~J1833$-$1034.
Spectral fittings were performed with the X-ray Spectral Fitting Package (XSPEC) version 12.9.0u \citep{Arnaud1996} with the Cash statistics \citep{Cash1979}.
We did not rebin the spectra since the Cash statistics can deal with low-count bins as opposed to the $\chi^2$ fitting method. 
We generated redistribution matrix files for the SXS and SXI with \texttt{sxsmkrmf} and \texttt{sxirmf}, respectively. 
We ran \texttt{aharfgen} \citep{Yaqoob2017} to generate ancillary response files for the SXS and SXI and and response files for the HXI. 
Since G21.5$-$0.9 has a faint diffuse extended halo out to $\sim140^{\prime\prime}$ from the pulsar (e.g., \cite{Matheson2005}), we generated the response files by inputing a Chandra image (0.5--10.0~keV) to  \texttt{aharfgen} to take into account the spatial extent.
Note however that whether the assumed source type is ``extended'' or ``point-like'',  our spectral analysis results are unaffected. 
The background spectrum for the SXI is extracted from a source-free region of the on-axis segment (CCD2CD).
Off-source spectra are used for the HXI backgrounds as well. 

\section{Wide-band Spectroscopy}
\subsection{Analysis}

\begin{figure}[htb]
 \begin{center}
  \FigureFile(8cm,){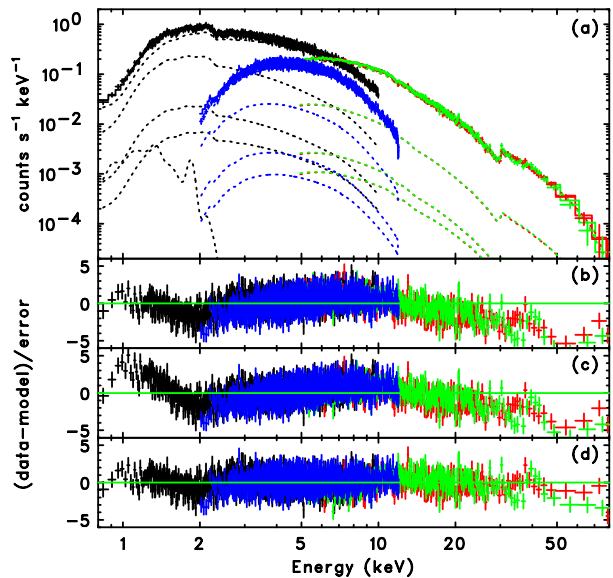} 
 \end{center}
\caption{Wide-band spectra of G21.5$-$0.9 obtained with the SXI (black; 0.8--10.0~keV), HXI1 (red; 5.0--80.0~keV), HXI2 (green; 5.0--80.0~keV) and the SXS (blue; 2.0--12.0~keV). The data is rebinned only for plotting purposes. The best-fit model (composite$+$Broken PL; see text and table~\ref{tab:parameters}) is overlaid with the solid lines in panel (a). The dotted lines indicate all the additive components in the model. Panels (b), (c), and (d) show residuals from the single PL, composite$+$PL and composite$+$Broken PL models, respectively. }\label{fig:spectrum}
\end{figure}

Figure~\ref{fig:spectrum} (a) shows the background-subtracted spectra of G21.5$-$0.9 (0.8--10.0~keV for the SXI, 5.0--80.0~keV for the HXI and 2.0--12.0~keV for the SXS).
The featureless spectral shape already suggests  that the emission is dominated by non-thermal X-ray emission, as reported by previous X-ray studies \citep{Slane2000, Safi-Harb2001, Warwick2001, Bocchino2005, Matheson2010, Tsujimoto2011, Nynka2014}.
 In order to fit the SXS, SXI and HXI data, 
 we first attempted a single power law (hereafter, single PL) modified by interstellar absorption using the Tuebingen--Boulder ISM absorption (TBabs in XSPEC; \cite{Wilms2000}).
 We find that while this model fits well the spectra up to $\sim10$~keV, giving a photon index of $\sim2.0$, it overpredicts the emission in the HXI band, suggesting a spectral break.
 The residuals and the fitting parameters are shown in figure~\ref{fig:spectrum} (b) and table~\ref{tab:parameters}, respectively.
When fitting the HXI data alone with the column density frozen to its best fit value from the broadband fit, we find a steeper photon index of $\sim2.2$, confirming our conclusion above.

Guided by the most recent spatially resolved Chandra studies of this source (\cite{Matheson2010, Guest2017}; see also \cite{Bocchino2005} for the XMM-Newton study) showing that the spectrum steepens away from the source and has some weak thermal X-ray emission from the northern knot, we used a ``composite'' model that accounts for the emission from all but the power-law emission from the PWN  (as observed with Chandra, \cite{Guest2017}). 
We define the model ``composite$+$PL'' as multiple components from the pulsar, the extended halo and the limb,  a weak, thermal soft  ($kT_{\rm e}\sim0.15$~keV) component from the northern knot, represented by a non-equilibrium ionization model (vpshock in XSPEC;  \cite{Borkowski2001}) plus a power-law component from the PWN (the most dominant component).
We note here that the SXS is not sensitive to the localized thermal component 
due to the limited sensitivity below $\sim2$~keV and the lack of spatial resolution to extract the thermal knots.
We also note that the blackbody thermal component from the pulsar, PSR~J1833$-$1034, reported by \citet{Matheson2010} is not significant and contributes with a negligible fraction to the spectrum of the SNR obtained with Hitomi.
As shown in figure~\ref{fig:spectrum} (c), we find that the model (composite$+$PL) is sufficient to explain the SXS data.
The model, however, underpredicts or overpredicts the soft and hard X-ray emissions detected with the SXI and HXI, respectively.
The result again clearly shows  negative residuals at $>10$~keV, which suggests that a steeper power-law slope is required by the HXI data, as claimed by recent studies obtained 
in the hard X-ray band \citep{Tsujimoto2011, Nynka2014}.
The best-fit results for the composite$+$PL model are displayed in table~\ref{tab:parameters}. 

\begin{table*}[tb]
  \caption{Spectral Fitting Results of the Hitomi G21.5$-$0.9 Data.}\label{tab:parameters}
  \begin{center}
    \begin{tabular}{lccc}
\hline
& \multicolumn{3}{c}{Model}\\
\cline{2-4}
Parameter & single PL &  composite$+$PL& composite$+$Broken PL \\
\hline	   						    		      			  						
$N\rm{_H}$ (10$^{22}$~cm$^{-2}$) & $3.50\pm0.03$  & $3.64\pm0.02$  & $3.22\pm0.03$ \\
$\Gamma_{1}$ &$2.03\pm0.01$&$2.01\pm0.01$&$1.74\pm0.02$\\
$\Gamma_{2}$ &---&--- &$2.14\pm0.01$\\
$E\rm_{break}$ (keV)&---&---&$7.1\pm0.3$\\
$F\rm_{X, soft}$\footnotemark[$*$] (10$^{-11}$~erg~s$^{-1}$~cm$^{-2}$) & $3.39\pm0.04$&$2.88\pm0.03$   &$4.80\pm0.02$\\
$F\rm_{X, hard}$\footnotemark[$\dagger$] (10$^{-11}$~erg~s$^{-1}$~cm$^{-2}$) & $4.96\pm0.04$&$4.92\pm0.04$ &$4.54\pm0.04$\\
\hline
C-statistics (using 23035 PHA bins) /d.o.f. & 25447.06/23030 &25380.67/23029  & 24228.18/23027\\
\hline
    \end{tabular}
  \end{center}
    \begin{tabnote}
    The errors are 90\% confidence level.\\
    \footnotemark[$*$] Intrinsic flux in the 2.0--8.0~keV range for the SXI and SXS. \\
    \footnotemark[$\dagger$] Intrinsic flux in the 15.0--50.0~keV range for the HXI. 
    \end{tabnote}
\end{table*}

We subsequently replaced the power-law model component representing the PWN with a broken power-law model (composite$+$Broken PL) to reproduce the spectral break.
The result and residuals are presented in figures~\ref{fig:spectrum} (a) and (d), respectively.
The model (composite$+$Broken PL) reduces the large residuals at $>10$~keV seen in figure~\ref{fig:spectrum} (c).
As shown in table~\ref{tab:parameters}, this model fits the spectra with photon indices of $\Gamma_1 = 1.74 \pm 0.02$ for the soft band and $\Gamma_2 = 2.14 \pm 0.02$ for the hard band, giving a break energy $E\rm_{break}=7.1\pm0.3$~keV.
We note that the best-fit column density of $N\rm{_H} = (3.2 \pm 0.03)\times10^{22}$~cm$^{-2}$ is lower than those obtained by \citet{Matheson2010} and previous Chandra and XMM-Newton studies. 
This is mainly due to the difference of the abundance tables used in the spectral fittings. 
We use here the updated abundance table  \citep{Wilms2000} whereas most previous X-ray studies used the abundances given by \citet{Anders1989}. 
The choice, however, does not affect the other spectral parameters such as the photon indices or the break energy.

\subsection{Origin of Spectral Break at $\sim7$~keV}\label{sec:Dspec}

We know from previous Chandra X-ray studies that the spectral index for G21.5$-$0.9 steepens away gradually from the PSR~J1833$-$1034 as we go out to the limb of the SNR \citep{Matheson2005}.
Here we demonstrate that the spectral softening or break required for fitting the HXI spectrum of the SNR cannot be due to this spatially-varying photon index;
that is, the addition of the different power law components does not reproduce the spectrum observed with the Hitomi data.
This conclusion was similarly reached by the NuSTAR study \citep{Nynka2014}.

\begin{figure}[ht]
 \begin{center}
 \FigureFile(7.8cm,){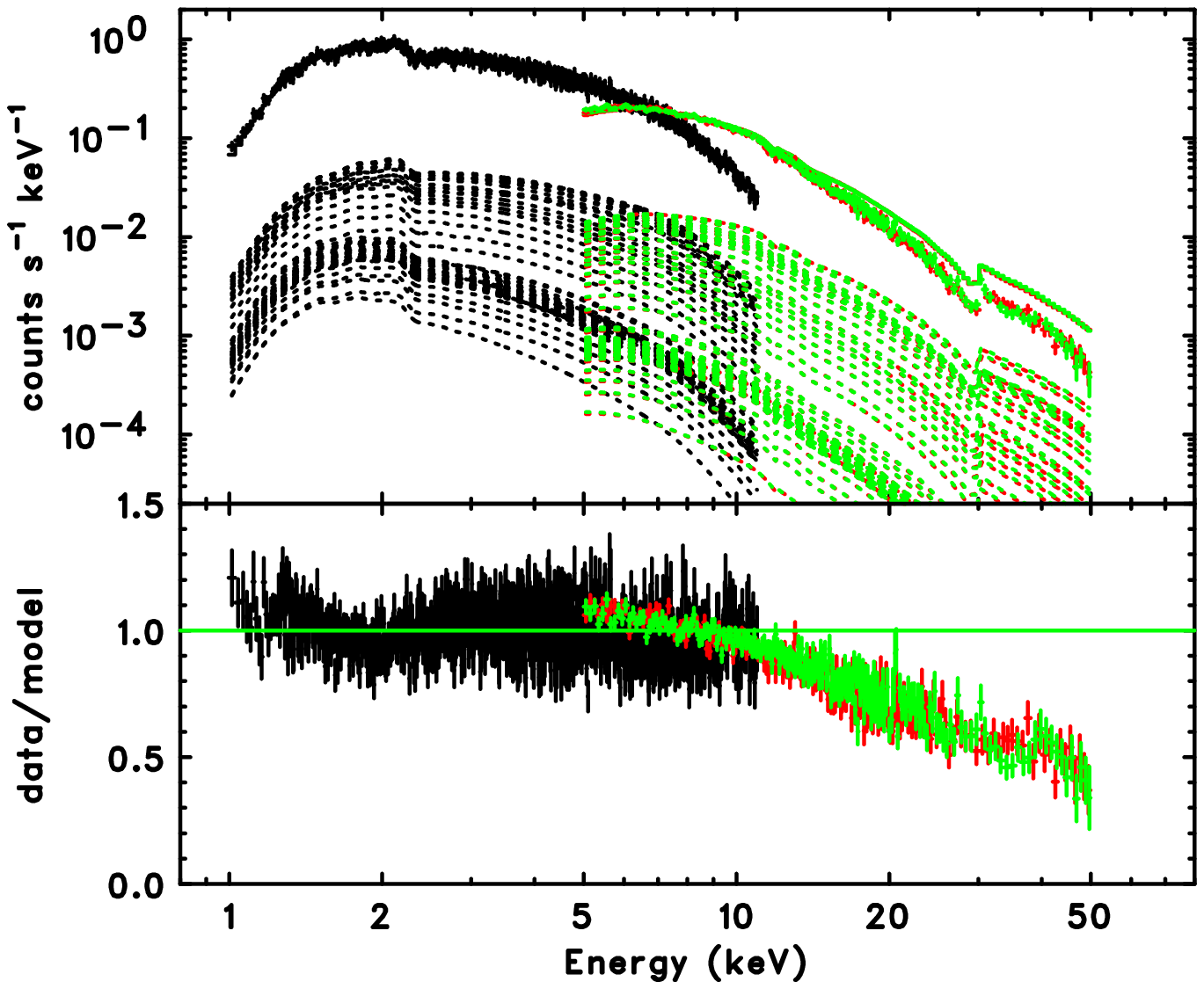}
  \end{center}
\caption{The SXI (black) and HXI1/2 data (red and green) fitted with the composite$+$PL model accounting for the spatially resolved spectroscopic study of the SNR with Chandra (\cite{Guest2017}; see also \cite{Matheson2005}).
The individual components contributing to the fitted spectrum are shown as dashed lines. The bottom panel shows the data-to-model ratios  
and illustrates that this model does not reproduce the spectral shape obtained with the HXI. }\label{fig:compositechandramodel}
\end{figure}

To that end, we construct a composite power-law model consisting of spatially resolved spectra of 50 regions obtained with all Chandra data acquired to date
(\cite{Guest2017}; see also \cite{Matheson2005}). The model accounts for the small-scale regions extending from the pulsar out to the SNR limb
and consists of power-law model components with an index steepening from $\sim1.5$ at the pulsar to $\sim2.6$ in the outermost region.
Fitting this composite model to the Hitomi spectra clearly shows that the model does not fit the HXI data, as shown in figure~\ref{fig:compositechandramodel}.

We have to consider possible mechanisms to make the spectral break other than the spatial variation of the synchrotron radiation. 
Let us discuss this in the context of a multi-wavelength study using data from radio up to TeV gamma rays including the Hitomi data. 
Many authors have been trying to reproduce spectral energy distributions of PWNe such as the Crab nebula and G21.5$-$0.9 in the literature (e.g., \cite{Atoyan1996,Zhang2008,Tanaka2010,Tanaka2011,Martin2012,Torres2014}). 
In what follows, we calculate emission models for G21.5$-$0.5 based on the one-zone model by \citet{Tanaka2010} and \citet{Tanaka2011}. 

The PWN is assumed to be a uniform sphere with a radius of $R_{\rm pwn}$ expanding with a constant velocity $v_{\rm pwn}$ (i.e., $R_{\rm pwn} = v_{\rm pwn}t$). 
The spin-down power of the central pulsar is expressed as 
\begin{eqnarray}
L_{\rm sd}(t) = L_{{\rm sd}0} \left( 1 + \frac{t}{\tau_0} \right)^{-\frac{n+1}{n-1}}, \label{eq:spindown}
\end{eqnarray}
where $L_{{\rm sd}0}$, $\tau_0$, and $n$ are the initial spin-down luminosity, the initial spin-down timescale, and the breaking index, respectively. 
The spin-down luminosity is finally converted either to kinetic power of relativistic positrons and electrons (we refer to simply as electrons hereafter) 
$L_e$ or into magnetic power $L_B$ in the PWN region. 
The ratio of the two channels is determined by the temporally and spatially constant parameter $\eta$ ($0 \leq \eta \leq 1$) as
\begin{eqnarray}
L_e(t) &=& (1- \eta) L_{\rm sd}(t), \label{eq:Le}\\
L_B(t) &=& \eta L_{\rm sd}(t). 
\end{eqnarray}
Electrons are injected to the PWN with a broken power-law spectrum:  
\begin{eqnarray}
Q(E,t) = \left\{ \begin{array}{ll}
    Q_0(t) (E/E_b)^{-p_1} & (E_{\rm min} \leq E < E_{\rm b}) \\
    Q_0(t) (E/E_b)^{-p_2} & (E_{\rm b} \leq E \leq E_{\rm max}) \\
    0 & ({\rm otherwise}), 
  \end{array} \right.
\end{eqnarray}
where $E$ denotes the kinetic energy of electrons and $E_b$ is the break energy. 
The normalization $Q_0(t)$ can be obtained by substituting 
\begin{eqnarray}
L_e(t) = \int^{E_{\rm max}}_{E_{\rm min}}EQ(E,t) \, dE
\end{eqnarray}
into equation (\ref{eq:Le}). 
The magnetic energy conservation,  
\begin{eqnarray}
\frac{4 \pi}{3} [R_{\rm PWN}(t)]^3 \frac{[B(t)]^2}{8 \pi} = \int^{t}_0 \eta L(t^\prime)\, dt^\prime,
\end{eqnarray}
together with equation (\ref{eq:spindown}) yields the magnetic field strength 
\begin{eqnarray}
B(t) = \sqrt{\frac{3(n-1)\eta L_{{\rm sd}0} \tau_0}{[R_{\rm PWN}(t)]^3} \left[ 1 - \left( 1 + \frac{t}{\tau_0}\right)^{-\frac{2}{n-1}}\right]}. 
\end{eqnarray}

The electron spectrum at time $t$ is obtained by solving the Fokker-Planck equation
\begin{eqnarray}
\frac{\partial N(E, t)}{\partial t} = \frac{\partial}{\partial E} [b(E,t)\ N(E, t)] + Q(E, t) 
\end{eqnarray}
for $N(E,t)$, where $b(E,t)$ is the energy loss rate of electrons. 
We consider energy losses by synchrotron, IC, and adiabatic expansion of the PWN. 
We then calculate synchrotron and IC  radiation spectra from the electrons with the spectrum 
$N(E,t_{\rm age})$, where $t_{\rm age}$ is the age of the pulsar. 
In the calculation of the synchrotron spectrum, we assume that the magnetic field line directions are randomly distributed, 
and use the analytical formula for the synchrotron spectrum from a single electron by \citet{Zirakashvili2007}. 
We consider isotropic radiation fields for IC, and calculate the spectrum by using the expression 
given by \citet{Jones1968}. 
The radiation fields spectra are taken from the model implemented in GALPROP \citep{Porter2006}, which 
includes the cosmic microwave background, optical radiation from stars, and infrared radiation due to reemission of the 
optical component by dust. 
 
\begin{figure}[t]
 \begin{center}
 \FigureFile(7.5cm,){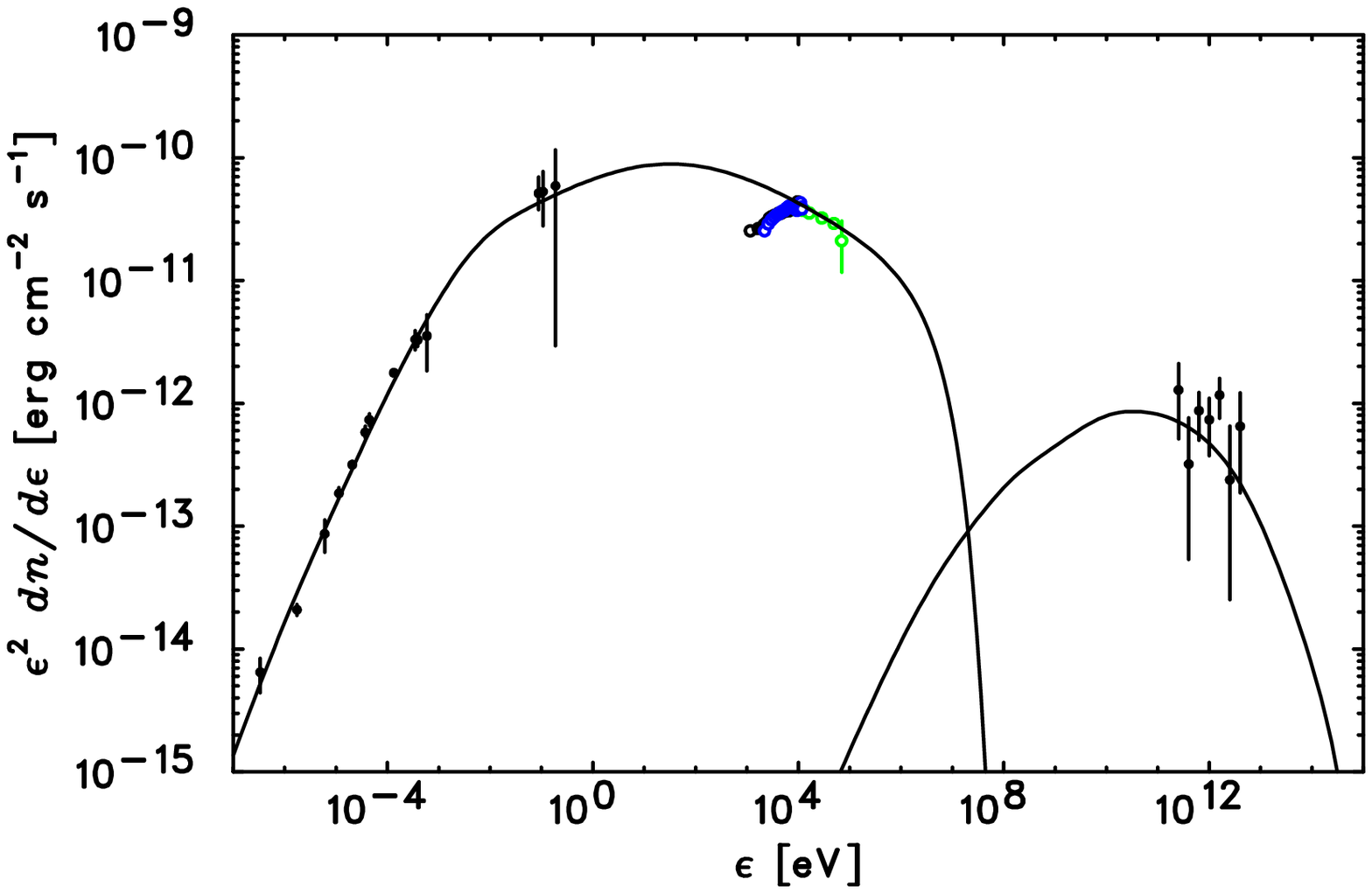}
\ \end{center}
\caption{Spectral energy distribution of G21.5$-$0.9 with the Case 1 model whose parameters are 
summarized in table~\ref{tab:sed_param}. The black, blue, and green data points in the X-ray band are from the SXI, SXS, and HXI, respectively. 
The data from HXI1 and HXI2 are co-added for display purpose. The radio data points are taken from \citet{Wilson1976}, \citet{Becker1975}, \citet{Morsi1987}, and \citet{Salter1989}
The infrared data are obtained with the Infrared Space Observatory by \citet{Gallant1999}. The H.E.S.S. data points in the TeV gamma-ray band are by \citet{Djannati-Atai2008}.}\label{fig:sed_case1}
\end{figure}

We first tried fitting the overall shape of the multi-wavelength spectrum of G21.5$-$0.9 (Case 1). 
Figure~\ref{fig:sed_case1} shows the result of the calculation plotted with the data in the radio, infrared, X-ray, and TeV gamma-ray bands. 
In the calculation, we assumed 4.7~kpc as the distance to the PWN \citep{Camilo2006}. 
Referring to \citet{Bietenholz2008}, we assumed the expansion velocity of the PWN and the age of the pulsar to be 
$v_{\rm pwn} = 910~{\rm km}~{\rm s}^{-1}$ and $t_{\rm age} = 870~{\rm yr}$, respectively.  
Since the second derivative of the pulsar period has not been measured, we simply assumed $n = 3$, which 
corresponds to spin-down via magnetic dipole radiation. 
The rotation period $P$ and period derivative $\dot{P}$ of PSR~J1833$-$1034 are taken from \citet{Camilo2006} as 
$P = 61.9~{\rm ms}$ and $\dot{P} = 2.02 \times 10^{-13}$, which are used to obtain $\tau_0$ and $L_{{\rm sd}0}\tau_0$ 
as 
\begin{eqnarray}
\tau_0 &=& \frac{P}{(n-1)\dot{P}} - t_{\rm age} = 4.0~{\rm kyr} \\
P_0 &=& P \left( 1 + \frac{t_{\rm age}}{\tau_0} \right)^{-\frac{1}{n-1}} = 56~{\rm ms} \\
L_ {{\rm sd}0} \tau_0 &=& \frac{I}{(n-1)\tau_0}\left(\frac{2\pi}{P_0}\right)^2 = 6.3 \times 10^{48}~{\rm erg}.   
\end{eqnarray}
Here $P_0$ is the initial pulsar period, and $I$ is pulsar's moment of inertia for which we assumed $10^{45}~{\rm g}~{\rm cm}^2$. 
The parameters are similar to those of Model~1 by \citet{Tanaka2011}. 
Although the model fits well the radio, infrared, and gamma-ray data points, it fails to fit the Hitomi spectra particularly in the 
soft X-ray band below the break at $7~{\rm keV}$.  

\begin{figure}[!ht]
 \begin{center}
 \FigureFile(7.5cm,){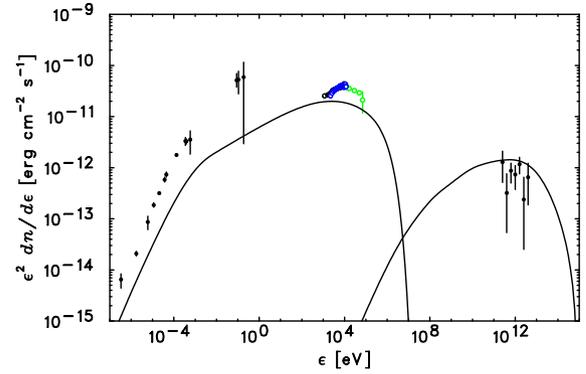}
\ \end{center}
\caption{Same as figure~\ref{fig:sed_case1} but with the Case 2 model curves.}\label{fig:sed_case2}
\end{figure}

One of the possible mechanisms to make the X-ray spectral break is synchrotron cooling. 
In the model presented in figure~\ref{fig:sed_case1}, the synchrotron cooling break appears at $\sim10^2~{\rm eV}$. 
Since the synchrotron cooling break energy is roughly proportional to $B^{-3}$, we need to have 
a weaker magnetic field and thus smaller $\eta$ to move the break toward a higher energy up to $7~{\rm keV}$ 
at which we found the break. 
In figure~\ref{fig:sed_case2}, we plot model curves for which we assumed smaller $\eta$ so that the synchrotron break 
coincides with the observed break (Case 2). 
The parameters are summarized in table~\ref{tab:sed_param}. 
Smaller $\eta$ results in a lower synchrotron-to-IC flux ratio, which contradicts the data.  
In addition, the model predicts a smaller spectral slope change at the break than the Hitomi data. 
The assumption about the magnetic field evolution in principle can affect the results. 
Several authors (e.g., \cite{Zhang2008,Torres2014}) indeed considered different magnetic field evolution models. 
The situation, however, would not be drastically improved even if we adopt their assumptions. 

\begin{table*}[tb]
\caption{Parameters for model calculations.}\label{tab:sed_param}
\begin{center}
\begin{tabular}{ccccccc}
\hline
 & $\eta$ & $E_{\rm min}$ & $E_{\rm b}$  &  $E_{\rm max}$  & $p_1$ & $p_2$\\
 \hline
Case 1& $2.0 \times 10^{-2}$ & 0.5~GeV & 50~GeV & 1~PeV & 1.0 & 2.5\\
Case 2 &  $1.0 \times 10^{-3}$ & 0.5~GeV  & 50~GeV & 1~PeV & 1.0 & 2.5 \\
\hline
\end{tabular}
\end{center}
\end{table*}

Instead of synchrotron cooling, another break in the electron injection spectrum might be able to explain the break we observed. 
This scenario, however, would not be feasible at least with a one-zone model. 
As demonstrated by the Case 1 model shown in figure~\ref{fig:sed_case1}, the parameter $\eta$ should be $\sim 10^{-2}$ to account for the observed synchrotron-to-IC ratio. 
In this case, the synchrotron cooling break inevitably appears at an energy below the X-ray band, which leads to a softer X-ray spectrum.  
It is then difficult to reproduce the low-energy part of the Hitomi spectrum, i.e., the hard spectrum below the break with a photon index of $\Gamma_1 = 1.7$. 

It is likely that more complicated models are required to reproduce the observational data. 
We assumed a single electron population in an emitting region where physical parameters such as the magnetic field strength 
are uniform. 
In reality, electrons are transported from the termination shock of the PWN through advection and diffusion \citep{deJager2008, Tang2012, Vorster2013}. 
Higher energy electrons suffer from significant synchrotron cooling, which makes the electron spectrum spatially variable. 
The magnetic field should have spatial variation as well. 
X-rays would be emitted by electrons close to the termination shock where the magnetic field is relatively high 
while the radio-to-infrared radiation might be coming from a larger region. 
In this context, it is of interest to note that the radio and X-ray images presented by \citet{Matheson2005} suggest 
different morphologies. The X-ray emission appears more concentrated close to the pulsar compared with the radio image. 
It is also possible that radio-emitting  and X-ray-emitting electrons have different origins. 
\citet{TanakaAsano2017} proposed such a model (see also \cite{Ishizaki2017}).  
In their model, electrons responsible for X-rays are provided by the pulsar wind and are 
accelerated at the termination shock through the diffusive shock acceleration process. 
On the other hand, radio-emitting electrons are supplied, for example,  by supernova ejecta, and are stochastically accelerated by 
turbulence inside a PWN. 
Such models could reproduce the complex synchrotron shape that the Hitomi result revealed.

\section{Search for Lines}\label{sec:STLines}
\subsection{Analysis}
We performed a blind search of emission and absorption lines from the SXS spectrum. 
We focus on narrow lines in the 2--10~keV band. 
The bandpass is limited by the attenuation by the closed gate-valve below 2 keV and the photon statistics above 10~keV. 
Features with a width up to 1280~km~s$^{-1}$ were searched. 
A search for weak broad features is strongly coupled with the exact shape of the continuum, details of which are hampered by the incomplete calibration of the effective area of the SXS \citep{Tsujimoto2017}.

We took the same approach as for the Crab nebula \citep{HitomiCrab2017}, in which we fitted the spectrum locally and added a single Gaussian model with a fixed trial energy and width. 
The trial energies are from 2 to 10~keV with a 0.5~eV step and the width are 0, 20, 40, 80, 160, 320, 640, and 1280~km~s$^{-1}$. The power-law model was used for the local continuum fitting in an energy range 3--20 $\sigma (E)$ on both sides of the trial energy $E$, in which $\sigma (E)$ is the quadrature sum of the trial width and the line spread function width. 
The significance of the detection was assessed as
\begin{equation}
 \sigma = \frac{N_{\mathrm{line}}}{\sqrt{\Delta {N_{\mathrm{line}}^2} +
  (N_{\mathrm{line}} \Delta I_{\mathrm{cont}}/I_{\mathrm{cont}})^2}}\label{e01},
\end{equation}
in which $N_{\mathrm{line}}$ and $\Delta N_{\mathrm{line}}$ are the best-fit and 1
$\sigma$ statistical uncertainty of the line normalization in the unit of
s$^{-1}$~cm$^{-2}$, whereas $I_{\mathrm{line}}$ and $\Delta I_{\mathrm{line}}$ are those
of the continuum intensity in the unit of s$^{-1}$~cm$^{-2}$~keV$^{-1}$ at the line
energy. Positive values indicate emission, whereas negative values indicate absorption.

\begin{figure}
 \begin{center}
  \FigureFile(8cm,){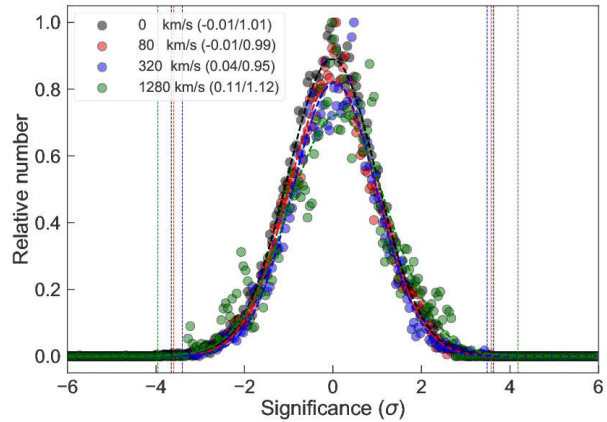}
 \end{center}
 \caption{Distribution of significance (equation~\ref{e01}) for several selected trial widths
 in different colors. The distribution is fitted by a single Gaussian model, and its
 best-fit parameter is shown in the legend as (center, width). The horizontal dotted
 lines indicate the significance at which the upper or lower probability is 0.1\%
 assuming the best-fit Gaussian distribution.} \label{f02}
\end{figure}

\begin{figure*}
 \begin{center}
 \FigureFile(8cm,){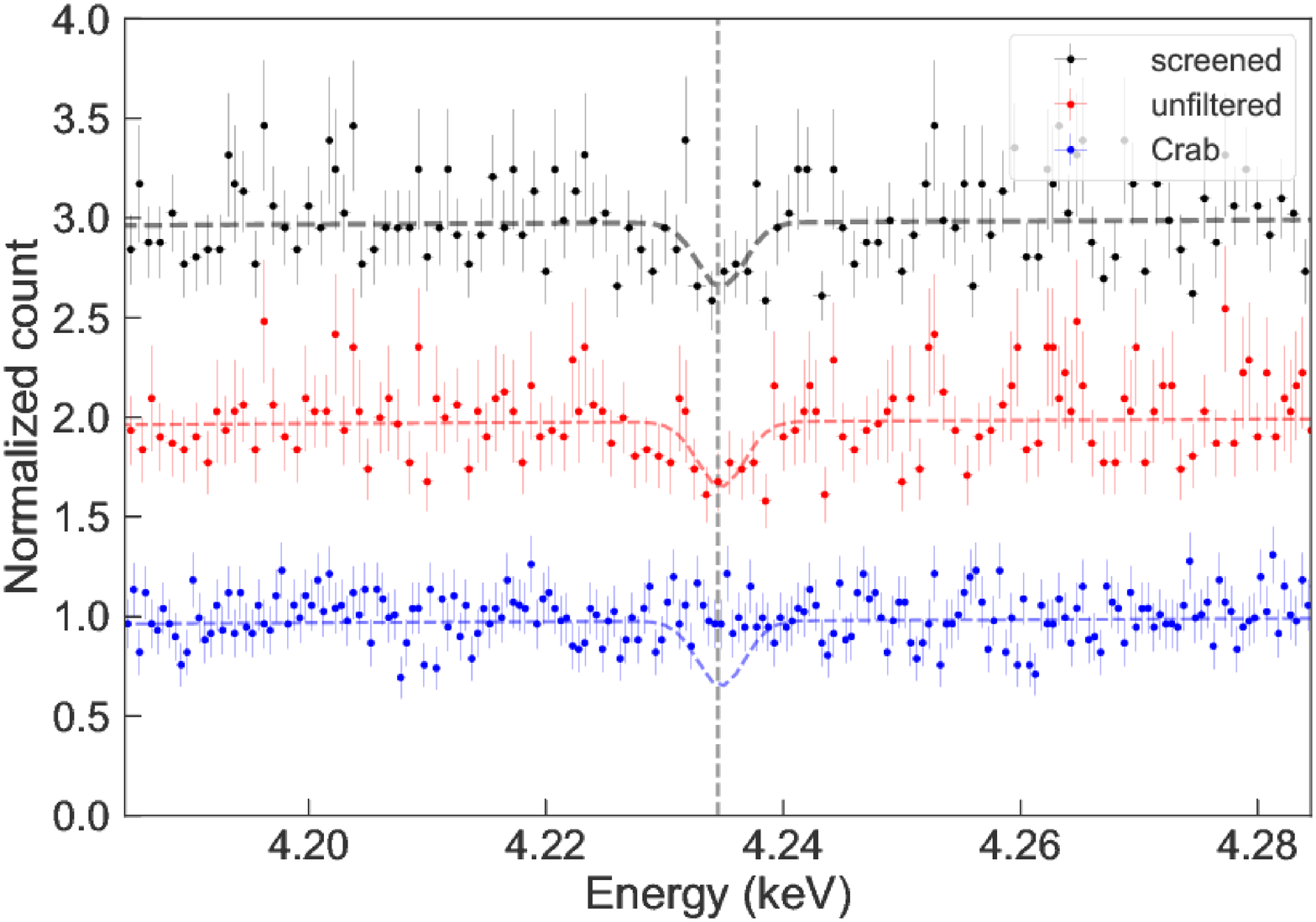}
 \FigureFile(8cm,){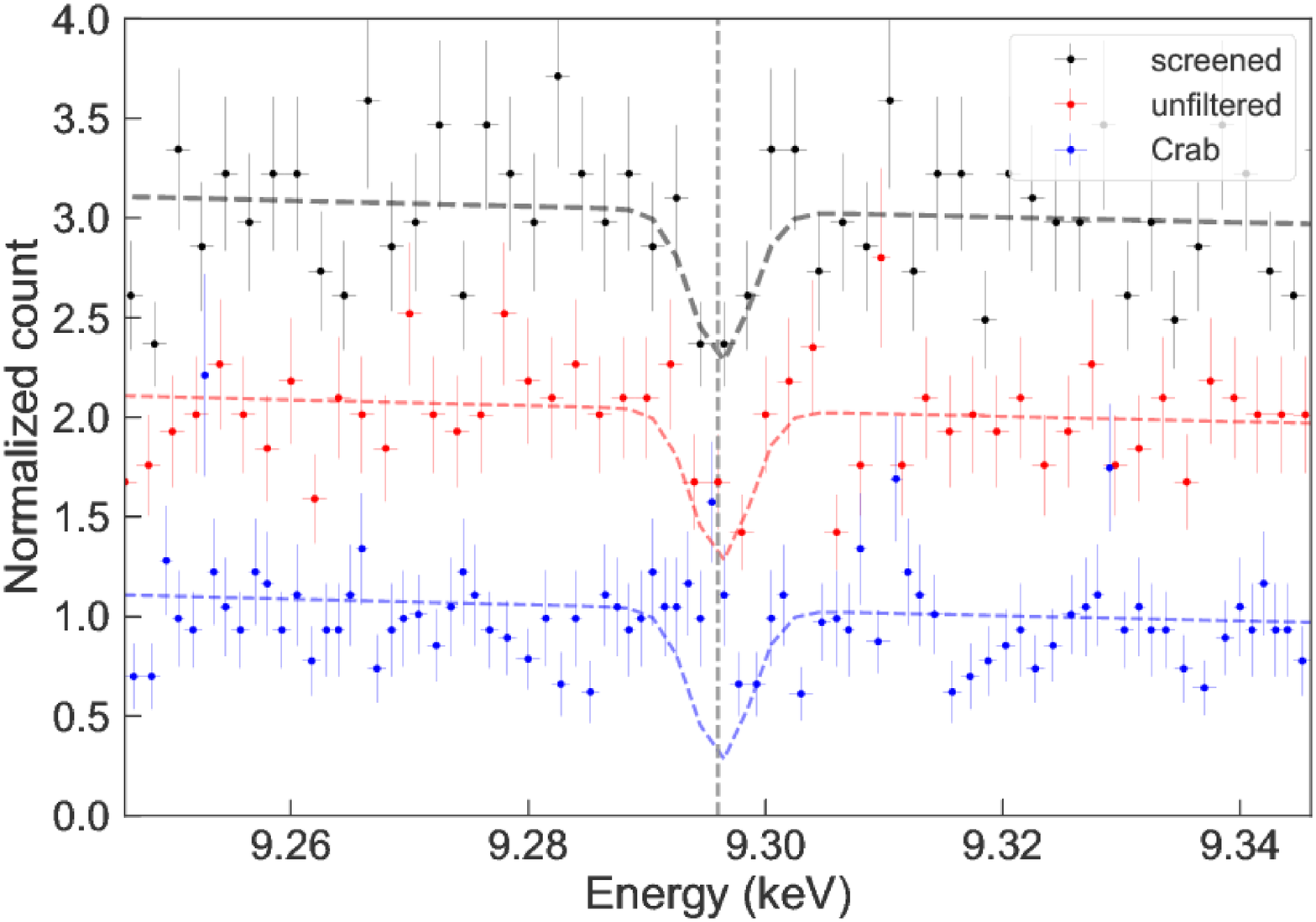}
 \end{center}
 \caption{Background-unsubtracted spectra at two energies (4.2345~keV and 9.296~keV for the
 left and right panels, respectively). Black, red, and blue respectively show the
 background-unsubtracted spectrum for the screened G21.5$-$0.9, the unfiltered G21.5$-$0.9, and the
 screened Crab data, which are normalized and offset to have a mean at 3.0, 2.0, and
 1.0. The black dotted curve is the best-fit continuum plus Gaussian model for a
 velocity width of 0~km~s$^{-1}$. The red and blue curves are the same model with a different
 offset to match with the comparison data.}  \label{f01}
\end{figure*}

Figure~\ref{f02} shows the distribution of significance for some selected trial widths. The
distribution of significances is well fitted by a simple Gaussian distribution. Assuming
that it is indeed a single Gaussian distribution, 
we set the detection limit such that, on both sides, there is less than 0.01 false positive for the number of trials.
There are nine trial absorption lines that lie in the tail of the distribution with significance of the  with deviations of 3.65~$\sigma$. 
All of these lines are either at 4.2345~keV or 9.296~keV. We show the fits to the two
most significant ones in figure~\ref{f01}. These modeled absorption
lines yield an equivalent width of $-2.3\pm0.8$~eV and  velocity widths of 50--400~km~s$^{-1}$ for 4.2345~keV and $-4.9\pm2.2$~eV and $<$89~km~s$^{-1}$ for 9.296~keV.
The results are summarized in table~\ref{tab:abs_lines}.
 
\begin{table*}[tb]
\caption{Parameters for detected absorption lines.}\label{tab:abs_lines}
\begin{center}
\begin{tabular}{ccccc}
\hline
  Line Centroid (keV) & Equivalent Width (eV) & Velocity Width (km~s$^{-1}$)&  Significance ($\sigma$) \\
 \hline
4.2345 & $-2.3\pm0.8$ & 50--400 & 3.65 \\
9.296  &  $-4.9\pm2.2$ & $<$89  & 3.65 \\
\hline
\end{tabular}
\end{center}
\end{table*}

In figure~\ref{f01}, for comparison, we also plot the G21.5$-$0.9 spectrum made with unfiltered events and the Crab spectrum
with screened events. The former is intended to examine artifacts by event screening,
while the latter by the effective area calibration. For both energies, the absorption features
are not seen in the Crab data (and other Hitomi datasets), indicating that they are not instrumental features. The
features are seen both in the unfiltered and screened spectra, suggesting that they are
not due to the screening. 

\subsection{Possible Absorption Line Features}
The method described above using the SXS data revealed absorption features around 4.2345~keV and 9.296~keV.
Given that these lines are not present in other Hitomi data, including the Crab (an object similar in nature to G21.5$-$0.9), we propose an astrophysical origin.
However, we cannot identify these lines as there is no known strong atomic transitions in nearby energies even if we consider doppler effect due to the expansion.

One interpretation is electron cyclotron resonance scattering.
The absorption feature would then be at  
\begin{eqnarray}
E_{\rm c}=11.6~\left(\frac{B}{10^{12}~\rm{G}}\right)~{\rm keV} \sim42~\left(\frac{B}{3.6\times10^{12}~\rm{G}}\right)~\rm{keV}, 
\end{eqnarray}
for a surface dipole magnetic field strength of the pulsar $B=3.6\times10^{12}$~G, which is estimated from $P$ and $\dot{P}$. 
If interpreted as electron cyclotron features, the absorption features would be associated with lower magnetic fields of the order of 4$\times$10$^{11}$~G and 8 $\times$10$^{11}$~G for 4.2345~keV and 9.296~keV lines,  respectively.
In this case, the absorbing electrons would be located higher in the magnetosphere.
However the line features are not as broad as we expect for cyclotron absorption lines, and the ratio of their energies (given the precise values determined by the SXS) is not $1:2$, as would be expected from harmonics.
We therefore rule out the possibility of the electron cyclotron absorption lines.

Another potential origin is surface atomic lines from the strongly magnetized neutron star atmosphere, as predicted by calculations with a high-field multiconfigurational Hartree-Fock code \citep[and references therein]{Miller1991, Miller1992}. 
While absorption features (or emission lines in a few cases) have been reported from a range of isolated neutron stars, from the extremely high magnetic field objects like magnetars (e.g., \cite{Turolla2015}),
 to the extremely low magnetic field objects like the Central Compact Objects (e.g., \cite{Bignami2003}),
 to the X-ray Dim Isolated Neutron Stars \citep{Borghese2017}, 
to even an isolated `ordinary' rotation-powered pulsar \citep{Kargaltsev2012},
these lines are all either relatively broad, or if similarly narrow (e.g., as seen in XMM-Newton gratings spectra of isolated neutron stars, \cite{Hohle2012}), 
they are at much lower energies. Furthermore, the presence of the lines is controversial in some of these sources.
The SXS features reported here in G21.5$-$0.9 are the first such narrow 
lines found in the hard X-ray band and for a rotation-powered pulsar powering a PWN.

More recently, \citet{Rajagopal1997} and \citet{Mori2007} constructed models of magnetized atmospheres composed of Fe and mid-$Z$ elements, respectively. 
According to their calculations and simulated spectra, multiple absorption features appear in the energy range from  $\sim 0.1~\rm{keV}$ up to $\sim 10~\rm{keV}$. 
We note that if the atmosphere is dominated by O or Ne \citep{Mori2007}, a magnetic field strength of $B>10^{13}$~G is required to explain the observed line feature at the energy as high as 9.296~keV.
Given the magnetic field of PSR~J1833$-$1034, $B=3.6\times10^{12}$~G, we speculate that heavier elements may be dominant in its atmosphere (unless we are probing higher order strong multipoles).
This then suggests  fallback of supernova ejecta onto the neutron star surface.
While the pulsar powering G21.5$-$0.9 is believed to be an isolated pulsar, the possibility of fallback would be interesting in the light of PSR~J1833$-$1034 being likely the youngest
known pulsar in our Galaxy with a PWN age estimated at only 870~yr \citep{Bietenholz2008}.
It is however difficult to identify a specific element only from the two faint features.
The Thomson depth has a complicated structure and the resultant spectra show many absorption lines whose centroids highly depend on $B$ and the temperature of the atmosphere \citep{Mori2007}.

Lastly, another potential origin is absorption associated with its surroundings, noting that the PWN has a significant dust scattering halo. 
Again however, the line energies are much too high to be associated with an ISM component. 
The lack of detection of X-ray pulsations (section~\ref{sec:SCPulsation}) hampers a phase-resolved spectroscopic study which would help differentiate between an intrinsic-to-the-pulsar or ambient origin.
Future deep observations of PSR~J1833$-$1034 with a high-resolution spectrometer, as well as the detection of similarly narrow hard X-ray absorption features from other similar systems,
will help reveal the nature of these features, and may open a new window for studying the atmospheres or environment of isolated pulsars.

\section{Search for Coherent Pulsation}\label{sec:SCPulsation}

\begin{table*}
  \tbl{{\bf Timing Search Results for Each Setting}}{%
  \begin{tabular}{ccccllc}
  \hline
Instrument & Region & Energy band (keV) & count$^{*}$ & $\chi^2$/d.o.f.$^\dagger$ & pulse fraction (\%) $^\dagger$ & count s$^{-1}$$^\ddagger$\\
  \hline
HXI&  $8''$ circle &30--40&   168 & 4.5,3.8,3.0,2.5 & 24,26,30,35 & $<2.4\times 10^{-4}$\\
HXI& $8''$ circle &40--50&    90 & 4.5,3.9,3.0,2.5 & 31,34,39,41 & $<1.6\times 10^{-4}$\\
HXI& $8''$ circle &50--60&    28 & 4.7,4.0,3.1,2.5 & 41,42,41,41 & $<5.8\times 10^{-5}$\\
HXI& $8''$ circle &60--70&    10 & 4.4,3.8,3.0,2.5 & 42,42,42,42 & $<2.1\times 10^{-5}$\\
HXI& $70''$ circle &30--40&  2768 & 4.7,4.0,3.1,2.5 &  6, 7, 9,10 & $<1.1\times 10^{-3}$\\
HXI& $70''$ circle &40--50&  1218 & 4.5,3.9,3.0,2.5 & 10,11,13,14 & $<7.3\times 10^{-4}$\\
HXI& $70''$ circle &50--60&   628 & 4.6,3.9,3.1,2.5 & 13,15,18,20 & $<5.2\times 10^{-4}$\\
HXI& $70''$ circle &60--70&   370 & 4.6,3.9,3.1,2.5 & 17,19,22,25 & $<3.9\times 10^{-4}$\\
SGD&         ---           &20--30& 11766 & 4.5,3.9,3.0,2.5 &  3, 3, 4, 5 & $<1.7\times 10^{-3}$\\
SGD&          ---           &30--50& 12401 & 4.7,3.9,3.1,2.5 &  3, 3, 4, 5 & $<1.8\times 10^{-3}$\\
SGD&          ---          &50--100& 17069 & 4.5,3.9,3.0,2.5 &  2, 3, 3, 4 & $<2.0\times 10^{-3}$\\
SGD&          ---         &100--200& 14855 & 4.4,3.8,3.0,2.5 &  2, 3, 3, 4 & $<1.7\times 10^{-3}$\\
  \hline
  \end{tabular}}\label{table:g215_timing_search}
\begin{tabnote}
\end{tabnote}
$*$ Total number of events, including background.\\
$\dagger$ 5-$\sigma$ upper limit by searches in the 5, 7, 13, and 23 phase bins, respectively.\\
$\ddagger$ 5-$\sigma$ upper limit in count rate.
\end{table*}

\begin{figure}
 \begin{center}
   \FigureFile(7.5cm,){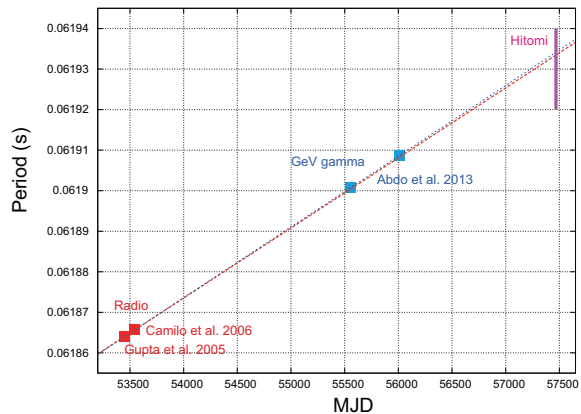}
 \end{center}
 \caption{Period measurements of PSR~J1833$-$1034 in the radio and GeV
gamma-ray band \citep{gupta2005,Camilo2006,abdo2013},
shown in red and cyan, respectively.
The search area with Hitomi is shown in magenta.}
\label{fig:TimingEvolution}
\end{figure}

We searched the HXI and SGD data for pulsed signals from the central pulsar PSR~J1833$-$1034.
Before analyzing the data, we estimated the expected period of the pulsar
during the Hitomi observation.
The measured $P$ in radio and GeV observations \citep{gupta2005,Camilo2006,abdo2013}
show straight linear increase with time
as shown in Figure~\ref{fig:TimingEvolution}.
The slope is consistent with $\dot{P}=2.2025(3)\times 10^{-13}$~s~s$^{-1}$,
the result of the most detailed observation \citep{Camilo2006}.
We thus decided to search $P$ in the range of
61.92--61.94~ms, and fixed $\dot{P}=2.2025\times 10^{-13}$~s~s$^{-1}$.

Extracting the HXI events, we tried two sizes of circular regions with  $8''$ and $70''$ radii centered at  
$\rm{(R.A.,~Dec.)=(18^{h}~33^{m}~33\fs8,~-10\arcdeg~34\arcmin~01\arcsec)}$
for  better signal-to-noise ratio for the pulsar against the PWN and the pulsar against the background, respectively.
In the extraction of the SGD events, the photo-absorption events were extracted following the method described in the appendix~2 in \citet{HitomiCrabGRP2017}.
We applied the barycentric correction on the arrival times of events using {\it barycen} for Hitomi \citep{HitomiTimeSystem2017}.
The timing searches were performed in each of the energy bands: 
20--30~keV, 30--40~keV, 40--50~keV, 50--60~keV, and 60--70~keV for the HXI, 
and 20--30~keV, 30--50~keV, 50--100~keV, and 100--200~keV for the SGD.
As a result, about 10--170 events were obtained per each energy band for the HXI smaller region, about 370--2,800 events for the HXI larger region, and about 12,000--17,000 events for the SGD.
We performed {\tt efserach} in HEAsoft 6.20 with the time resolution of 1~ns
on four sets of phase bin sizes (5, 7, 13, and 23 bins) 
with five different time origins (shifted by 0, 20\%, 40\%, 60\%, and 80\% of each phase-bin size)
and found no significant pulsation 
(i.e., the values of $\chi^2$/d.o.f. of trial-pulse profiles to the constant model are close to unity for all the trials).
We estimated the 5~$\sigma$ values of the $\chi^2$/d.o.f. on all the trials, as summarized in table \ref{table:g215_timing_search}.
In comparison of these $\chi^2$ values with the numerical simulations of possible pulses under the assumption that the pulse profiles have sinusoidal shapes in various amplitudes, the pulse fractions corresponding to the 5~$\sigma$ values of the $\chi^2$/d.o.f. were also estimated (table \ref{table:g215_timing_search}); the pulse fractions become similar values among various phase-bin settings although  $\chi^2$/d.o.f. varies by the settings.
The 5~$\sigma$ upper limit in the count rate in each energy band were also estimated in the table.
We also tried $Z^m$ analysis \citep{buccheri1983,brazier1994} for the same data set, in order to reduce high frequency noise.
Again, no significant pulsation was found.

\section{Summary}
While a standard pulsar wind theory of the Crab Nebula has been established by \citet{Kennel1984}, there are many evolution models proposed to generally describe the spectra of PWNe from radio to gamma rays.
G21.5$-$0.9 is a good example to investigate the emission mechanism in this context since the remnant is considered to be a prototype pulsar/PWN system in the early stage of the evolution (cf. \cite{Gaensler2006}).
We observed G21.5$-$0.9 with Hitomi on 2016 March 19--23 during the instrument commissioning and  verification phase of the satellite.
Thanks to their high sensitivity, wide band spectra obtained with the SXS, SXI and HXI onboard Hitomi revealed a detailed spectral feature in the range of 0.8--80~keV where a spectral break had been pointed out by previous studies \citep{Tsujimoto2011, Nynka2014}.
We constructed a ``composite'' spectral model accounting for all components of G21.5$-$0.9 to constrain the break energy of the central PWN.
Our results indicate that the PWN spectrum is reproduced by a broken power-law model with photon indices of $\Gamma_1=1.74\pm0.02$ and  $\Gamma_2=2.14\pm0.01$ below and above the 
break, respectively.
The break energy $E\rm_{break}$ is located at $7.1\pm0.3$~keV, which is significantly lower than that estimated from the NuSTAR spectra ($9.0^{+0.6}_{-0.4}$~keV in the 30\arcsec  \ inner region) by \citet{Nynka2014}.
We attempted to explain the SED from radio to TeV gamma rays with a spectral evolution model based on the work by \citet{Tanaka2010} and \citet{Tanaka2011}.
The overall shape of the multi-wavelength spectrum is well fitted by the model, whereas it fails to reproduce the Hitomi spectra particularly in the soft X-ray band below the break.
Our results require more complicated models considering, for example, stochastic acceleration (e.g., \cite{TanakaAsano2017}).
We also performed a timing analysis and a thermal line search of G21.5$-$0.9 with the Hitomi instruments: no significant pulsation was found from PSR~J1833$-$1034 with the HXI and SGD.
Two narrow absorption line features were detected at 3.65~$\sigma$ confidence at 4.2345~keV and 9.296~keV in the SXS spectrum. 
The observed absorption features reported here are not seen in the Crab data or other Hitomi datasets, suggesting that they are not an instrumental artifact. 
The nature of these features is not well understood, but their mere detection opens up a new area of research in the physics of plerions  and isolated pulsars and is a challenge to present-day models. 
It is highly surprising in that the spectrum of what was supposed to be a featureless calibration source shows significant unexpected spectral features. 
This indicates the power of the X-ray microcalorimeter for opening up a new discovery space in astrophysics.


\section*{Author Contributions}
H. Uchida, T. Tanaka, and S. Safi-Harb led the data analysis and draft preparation.
The wide-band spectroscopy was performed 
mainly by H. Uchida, T. Tanaka, S. Safi-Harb, Y. Maeda, N. Nakaniwa, and B. Guest.
The thermal line search was done by M. Tsujimoto and T. Sato.
Y. Terada took responsibility for the timing analysis with the help of H. Murakami.
A. Bamba coordinated the analysis tasks for each topic.
The paper was improved by J. P. Hughes, R. Mushotzky, and M. Sawada.

\bigskip
\begin{ack}

We thank D. A.~Smith and M. Kerr to giving us the detailed information on the Fermi LAT observations of PSR~J1833$-$1034.
We thank the support from the JSPS Core-to-Core Program.
We acknowledge all the JAXA members who have contributed to the ASTRO-H
(Hitomi) project.  All U.S. members gratefully acknowledge support
through the NASA Science Mission Directorate. Stanford and SLAC
members acknowledge support via DoE contract to SLAC National
Accelerator Laboratory DE-AC3-76SF00515.
Part of this work was performed under the auspices of the U.S. DoE by
LLNL under Contract DE-AC52-07NA27344.
Support from the European Space Agency is gratefully
acknowledged.
French members acknowledge support from CNES, the Centre
National d'\'{E}tudes Spatiales.
SRON is supported by NWO, the Netherlands Organization for Scientific Research.
Swiss team acknowledges support of the Swiss Secretariat for Education, Research and Innovation (SERI).
The Canadian Space Agency is acknowledged for the support of Canadian members.
We acknowledge support from
JSPS/MEXT KAKENHI grant numbers
JP15H00773, JP15H00785, JP15H02070, JP15H02090, JP15H03639, JP15H03641, JP15H03642, JP15H05438, JP15H06896, JP15K05107, JP15K17610, JP15K17657, JP16H00949, JP16H03983, JP16H06342, JP16J02333, JP16K05295, JP16K05296, JP16K05300, JP16K05309, JP16K13787, JP16K17667, JP16K17672, JP16K17673, JP17H02864, JP17K05393, JP21659292, JP23340055, JP23340071, JP23540280, JP24105007, JP24540232, JP25105516, JP25109004, JP25247028, JP25287042, JP25400236, JP25800119, JP26109506, JP26220703, JP26400228, JP26610047, and JP26800102.
The following NASA grants are acknowledged: NNX15AC76G, NNX15AE16G, NNX15AK71G, NNX15AU54G, NNX15AW94G, and NNG15PP48P to Eureka Scientific.
This work was partly supported by Leading Initiative for Excellent Young Researchers, MEXT, Japan, and also by the Research Fellowship of JSPS for Young Scientists.
H. Akamatsu acknowledges support of
NWO via Veni grant.
C. Done acknowledges STFC funding under grant ST/L00075X/1.
A. Fabian and C. Pinto acknowledge ERC Advanced Grant 340442.
P. Gandhi acknowledges JAXA International Top Young
Fellowship and UK Science and Technology Funding Council (STFC) grant
ST/J003697/2.
Y. Ichinohe and K. Nobukawa are supported by the Research Fellow of JSPS for Young Scientists.
N. Kawai is supported by the Grant-in-Aid for Scientific Research on Innovative Areas ``New Developments in Astrophysics Through Multi-Messenger Observations of Gravitational Wave Sources''.
S. Kitamoto is partially supported by the MEXT Supported Program for the Strategic Research Foundation at Private Universities, 2014-2018.
B. McNamara and S. Safi-Harb acknowledge support from NSERC.
T. Dotani, T. Takahashi, T. Tamagawa, M. Tsujimoto, and Y. Uchiyama acknowledge support from the Grant-in-Aid for Scientific Research on Innovative Areas ``Nuclear Matter in Neutron Stars Investigated by Experiments and Astronomical Observations''.
N. Werner is supported by the Lend\"ulet LP2016-11 grant from the Hungarian Academy of Sciences.
D. Wilkins is supported by NASA through Einstein Fellowship grant number PF6-170160, awarded by the Chandra X-ray Center, operated by the Smithsonian Astrophysical Observatory for NASA under contract NAS8-03060
We thank contributions by many companies, including
in particular, NEC, Mitsubishi Heavy Industries, Sumitomo Heavy
Industries, and Japan Aviation Electronics Industry.

Finally, we acknowledge strong support from the following engineers.
JAXA/ISAS: Chris Baluta, Nobutaka Bando, Atsushi Harayama, Kazuyuki Hirose, Kosei Ishimura, Naoko Iwata, Taro Kawano, Shigeo Kawasaki, Kenji Minesugi, Chikara Natsukari, Hiroyuki Ogawa, Mina Ogawa, Masayuki Ohta, Tsuyoshi Okazaki, Shin-ichiro Sakai, Yasuko Shibano, Maki Shida, Takanobu Shimada, Atsushi Wada, Takahiro Yamada; JAXA/TKSC: Atsushi Okamoto, Yoichi Sato, Keisuke Shinozaki, Hiroyuki Sugita; Chubu U: Yoshiharu Namba; Ehime U: Keiji Ogi; Kochi U of Technology: Tatsuro Kosaka; Miyazaki U: Yusuke Nishioka; Nagoya U: Housei Nagano; NASA/GSFC: Thomas Bialas, Kevin Boyce, Edgar Canavan, Michael DiPirro, Mark Kimball, Candace Masters, Daniel Mcguinness, Joseph Miko, Theodore Muench, James Pontius, Peter Shirron, Cynthia Simmons, Gary Sneiderman, Tomomi Watanabe; ADNET Systems: Michael Witthoeft, Kristin Rutkowski, Robert S. Hill, Joseph Eggen; Wyle Information Systems: Andrew Sargent, Michael Dutka;
Noqsi Aerospace Ltd: John Doty; Stanford U/KIPAC: Makoto Asai, Kirk Gilmore; ESA (Netherlands): Chris Jewell; SRON: Daniel Haas, Martin Frericks, Philippe Laubert, Paul Lowes; U of Geneva: Philipp Azzarello; CSA: Alex Koujelev, Franco Moroso.

\end{ack}



\begin{thebibliography}{}
\bibitem[Abdo et al.(2013)]{abdo2013}Abdo, A.~A., Ajello, M., Allafort, A., et al.\ 2013, \apjs, 208, 17
\bibitem[Altenhoff et al.(1970)]{Altenhoff1970} Altenhoff, W.~J., Downes, D., Goad, L., Maxwell, A., \& Rinehart, R.\ 1970, \aaps, 1, 319 
\bibitem[Anders \& Grevesse(1989)]{Anders1989} Anders, E., \& Grevesse, N.\ 1989, \gca, 53, 197 
\bibitem[Arnaud(1996)]{Arnaud1996} Arnaud, K.~A.\ 1996, Astronomical Data Analysis Software and Systems V, 101, 17 
\bibitem[Atoyan \& Aharonian(1996)]{Atoyan1996} Atoyan, A.~M., \& Aharonian, F.~A.\ 1996, \mnras, 278, 525 
\bibitem[Awaki et al.(2014)]{Awaki2014} Awaki, H., Kunieda, H., Ishida, M., et al.\ 2014, \ao, 53, 7664 
\bibitem[Bamba et al.(2010)]{bamba2010}Bamba, A., Mori, K., \& Shibata, S.\ 2010, \apj, 709, 507
\bibitem[Becker \& Kundu(1975)]{Becker1975} Becker, R.~H., \& Kundu, M.~R.\ 1975, \aj, 80, 679 
\bibitem[Becker \& Szymkowiak(1981)]{Becker1981} Becker, R.~H., \& Szymkowiak, A.~E.\ 1981, \apjl, 248, L23 
\bibitem[Bietenholz \& Bartel(2008)]{Bietenholz2008} Bietenholz, M.~F., \& Bartel, N.\ 2008, \mnras, 386, 1411 
\bibitem[Bignami et al.(2003)]{Bignami2003} Bignami, G.~F., Caraveo, P.~A., De Luca, A., \& Mereghetti, S.\ 2003, \nat, 423, 725 
\bibitem[Bocchino et al.(2005)]{Bocchino2005} Bocchino, F., van der Swaluw, E., Chevalier, R., \& Bandiera, R.\ 2005, \aap, 442, 539 
\bibitem[Borghese et al.(2017)]{Borghese2017} Borghese, A., Rea, N., Coti Zelati, F., et al.\ 2017, \mnras, 468, 2975 
\bibitem[Borkowski et al.(2001)]{Borkowski2001} Borkowski, K.~J., Lyerly, W.~J., \& Reynolds, S.~P.\ 2001, \apj, 548, 820 
\bibitem[Brazier(1994)]{brazier1994}Brazier, K.~T.~S.\ 1994, \mnras, 268, 709
\bibitem[Buccheri et al.(1983)]{buccheri1983}Buccheri, R., Bennett, K., Bignami, G.~F., et al.\ 1983, \aap, 128, 245 
\bibitem[Cash(1979)]{Cash1979} Cash, W.\ 1979, \apj, 228, 939 
\bibitem[Camilo et al.(2006)]{Camilo2006} Camilo, F., Ransom, S.~M., Gaensler, B.~M., et al.\ 2006, \apj, 637, 456 
\bibitem[de Jager et al.(2008)]{deJager2008} de Jager, O.~C., Ferreira, S.~E.~S., \& Djannati-Ata{\"i}, A.\ 2008, American Institute of Physics Conference Series, 1085, 199 
\bibitem[Djannati-Ata{\"i} et al.(2008)]{Djannati-Atai2008} Djannati-Ata{\"i}, A., deJager, O.~C., Terrier, R., Gallant, Y.~A., \& Hoppe, S.\ 2008, International Cosmic Ray Conference, 2, 823 
\bibitem[Fang \& Zhang(2010)]{Fang2010} Fang, J., \& Zhang, L.\ 2010, \aap, 515, A20 
\bibitem[Gaensler \& Slane(2006)]{Gaensler2006} Gaensler, B.~M., \& Slane, P.~O.\ 2006, \araa, 44, 17 
\bibitem[Gallant \& Tuffs(1999)]{Gallant1999} Gallant, Y.~A., \& Tuffs, R.~J.\ 1999, The Universe as Seen by ISO, 427, 313 
\bibitem[Guest \& Safi-Harb (2018)]{Guest2017}Guest B. \& Safi-Harb S. 2018, in preparation
\bibitem[Gupta et al.(2005)]{gupta2005}Gupta, Y., Mitra, D., Green, D.~A., \& Acharyya, A.\ 2005, Current Science, 89, 853 
\bibitem[Hagino et al.(2018)]{HitomiHXIOrbit2017}Hagino, K., et al.\ 2018, JATIS submitted
\bibitem[Hitomi Collaboration et al.(2018a)]{HitomiCrab2017} Hitomi Collaboration, Aharonian, F., Akamatsu, H., et al.\ 2017a, PASJ in press, DOI:10.1093/pasj/psx072 (arXiv:1707.00054)
\bibitem[Hitomi Collaboration et al.(2018b)]{HitomiCrabGRP2017} Hitomi Collaboration, Aharonian, F., Akamatsu, H., et al.\ 2017b, PASJ in press, DOI:10.1093/pasj/psx083 (arXiv:1707.08801)
\bibitem[Hohle et al.(2012)]{Hohle2012} Hohle, M.~M., Haberl, F., Vink, J., de Vries, C.~P., \& Neuh{\"a}user, R.\ 2012, \mnras, 419, 1525 
\bibitem[Hughes et al.(2000)]{hughes2000} Hughes, J.~P., Rakowski, C.~E., Burrows, D.~N., \& Slane, P.~O.\ 2000, \apjl, 528, L109 
\bibitem[Ishizaki et al.(2017)]{Ishizaki2017} Ishizaki, W., Tanaka, S.~J., Asano, K., \& Terasawa, T.\ 2017, \apj, 838, 142 
\bibitem[Jones(1968)]{Jones1968} Jones, F.~C.\ 1968, Physical Review, 167, 1159 
\bibitem[Kargaltsev \& Pavlov(2008)]{kargaltsev2008}Kargaltsev, O., \& Pavlov, G.~G.\ 2008, 40 Years of Pulsars: Millisecond Pulsars, Magnetars and More, 983, 171 
\bibitem[Kargaltsev et al.(2012)]{Kargaltsev2012} Kargaltsev, O., Durant, M., Misanovic, Z., \& Pavlov, G.~G.\ 2012, Science, 337, 946 
\bibitem[Kelley et al.(2016)]{Kelley2016} Kelley, R.~L., Akamatsu, H., Azzarello, P., et al.\ 2016, \procspie, 9905, 99050V 
\bibitem[Kennel \& Coroniti(1984)]{Kennel1984} Kennel, C.~F., \& Coroniti, F.~V.\ 1984, \apj, 283, 694 
\bibitem[Kirsch et al.(2005)]{Kirsch2005} Kirsch, M.~G., Briel, U.~G., Burrows, D., et al.\ 2005, \procspie, 5898, 22 
\bibitem[Matheson \& Safi-Harb(2005)]{Matheson2005} Matheson, H., \& Safi-Harb, S.\ 2005, Advances in Space Research, 35, 1099 
\bibitem[Matheson \& Safi-Harb(2010)]{Matheson2010} Matheson, H., \& Safi-Harb, S.\ 2010, \apj, 724, 572 
\bibitem[Matsumoto et al.(2017)]{HitomiHXT2017} Matsumoto, H., et al.\ 2017, JATIS submitted
\bibitem[Mart{\'{\i}}n et al.(2012)]{Martin2012} Mart{\'{\i}}n, J., Torres, D.~F., \& Rea, N.\ 2012, \mnras, 427, 415 
\bibitem[Meszaros \& Nagel(1985)]{Meszaros1985} Meszaros, P., \& Nagel, W.\ 1985, \apj, 298, 147
\bibitem[Miller \& Neuhauser(1991)]{Miller1991} Miller, M.~C., \& Neuhauser, D.\ 1991, \mnras, 253, 107 
\bibitem[Miller(1992)]{Miller1992} Miller, M.~C.\ 1992, \mnras, 255, 129 
\bibitem[Mori \& Ho(2007)]{Mori2007} Mori, K., \& Ho, W.~C.~G.\ 2007, \mnras, 377, 905 
\bibitem[Morsi \& Reich(1987)]{Morsi1987} Morsi, H.~W., \& Reich, W.\ 1987, \aaps, 69, 533 
\bibitem[Nakajima et al.(2018)]{Nakajima2017} Nakajima, H., Maeda, Y., Uchida, H., et al.\ 2018, PASJ in press, DOI:10.1093/pasj/psx116 (arXiv:1709.08829)
\bibitem[Nakazawa et al.(2018)]{HitomiHXI2017} Nakazawa, K., et al.\ 2018, JATIS submitted
\bibitem[Nynka et al.(2014)]{Nynka2014} Nynka, M., Hailey, C.~J., Reynolds, S.~P., et al.\ 2014, \apj, 789, 72 
\bibitem[Okajima et al.(2016)]{Okajima2016} Okajima, T., Soong, Y., Serlemitsos, P., et al.\ 2016, \procspie, 9905, 99050Z 
\bibitem[Pacini \& Salvati(1973)]{Pacini1973} Pacini, F., \& Salvati, M.\ 1973, \apj, 186, 249 
\bibitem[Porter et al.(2006)]{Porter2006} Porter, T.~A., Moskalenko, I.~V., \& Strong, A.~W.\ 2006, \apjl, 648, L29 
\bibitem[Rajagopal et al.(1997)]{Rajagopal1997} Rajagopal, M., Romani, R.~W., \& Miller, M.~C.\ 1997, \apj, 479, 347 
\bibitem[Rees \& Gunn(1974)]{Rees1974} Rees, M.~J., \& Gunn, J.~E.\ 1974, \mnras, 167, 1 
\bibitem[Reynolds \& Chevalier(1984)]{Reynolds1984} Reynolds, S.~P., \& Chevalier, R.~A.\ 1984, \apj, 278, 630 
\bibitem[Reynolds \& Keohane(1999)]{Reynolds1999} Reynolds, S.~P., \& Keohane, J.~W.\ 1999, \apj, 525, 368 
\bibitem[Safi-Harb et al.(2001)]{Safi-Harb2001} Safi-Harb, S., Harrus, I.~M., Petre, R., et al.\ 2001, \apj, 561, 308 
\bibitem[Salter et al.(1989)]{Salter1989} Salter, C.~J., Reynolds, S.~P., Hogg, D.~E., Payne, J.~M., \& Rhodes, P.~J.\ 1989, \apj, 338, 171 
\bibitem[Slane et al.(2000)]{Slane2000} Slane, P., Chen, Y., Schulz, N.~S., et al.\ 2000, \apjl, 533, L29 
\bibitem[Soong et al.(2014)]{Soong2014} Soong, Y., Okajima, T., Serlemitsos, P.~J., et al.\ 2014, \procspie, 9144, 914428 
\bibitem[Takahashi et al.(2016)]{Takahashi2014} Takahashi, T., Kokubun, M., Mitsuda, K., et al.\ 2016, \procspie, 9905, 99050U 
\bibitem[Tanaka et al.(2018)]{Tanaka2017} Tanaka, T., Uchida, H., Nakajima, H., et al.\ 2018, JATIS, in press (arXiv:1801.06932) 
\bibitem[Tanaka \& Takahara(2010)]{Tanaka2010} Tanaka, S.~J., \& Takahara, F.\ 2010, \apj, 715, 1248 
\bibitem[Tanaka \& Takahara(2011)]{Tanaka2011} Tanaka, S.~J., \& Takahara, F.\ 2011, \apj, 741, 40  
\bibitem[Tanaka \& Asano(2017)]{TanakaAsano2017} Tanaka, S.~J., \& Asano, K.\ 2017, \apj, 841, 78 
\bibitem[Tang \& Chevalier(2012)]{Tang2012} Tang, X., \& Chevalier, R.~A.\ 2012, \apj, 752, 83 
\bibitem[Terada et al.(2017)]{HitomiTimeSystem2017} Terada, Y., Yamaguchi, S., Sugimoto, S., et al.\ 2017, JATIS in press (arXiv:1712.01484)
\bibitem[Torres et al.(2014)]{Torres2014} Torres, D.~F., Cillis, A., Mart{\'{\i}}n, J., \& de O{\~n}a Wilhelmi, E.\ 2014, Journal of High Energy Astrophysics, 1, 31 
\bibitem[Tsujimoto et al.(2011)]{Tsujimoto2011} Tsujimoto, M., Guainazzi, M., Plucinsky, P.~P., et al.\ 2011, \aap, 525, A25 
\bibitem[Tsujimoto et al.(2018)]{Tsujimoto2017} Tsujimoto, M., Okajima, T., Eckart, M.~E., et al.\ 2018, PASJ in press, DOI:10.1093/pasj/psy008 (arXiv:1801.02104)
\bibitem[Turolla et al.(2015)]{Turolla2015} Turolla, R., Zane, S., \& Watts, A.~L.\ 2015, Reports on Progress in Physics, 78, 116901 
\bibitem[Vorster \& Moraal(2013)]{Vorster2013} Vorster, M.~J., \& Moraal, H.\ 2013, \apj, 765, 30 
\bibitem[Warwick et al.(2001)]{Warwick2001} Warwick, R.~S., Bernard, J.-P., Bocchino, F., et al.\ 2001, \aap, 365, L248 
\bibitem[Watanabe et al.(2016)]{Watanabe2016} Watanabe, S., Tajima, H., Fukazawa, Y., et al.\ 2016, \procspie, 9905, 990513 
\bibitem[Wilms et al.(2000)]{Wilms2000} Wilms, J., Allen, A., \& McCray, R.\ 2000, \apj, 542, 914 
\bibitem[Wilson \& Weiler(1976)]{Wilson1976} Wilson, A.~S., \& Weiler, K.~W.\ 1976, \aap, 53, 89 
\bibitem[Yaqoob et al.(2018)]{Yaqoob2017} Yaqoob, T., et al.\ 2018, JATIS submitted
\bibitem[Zhang et al.(2008)]{Zhang2008} Zhang, L., Chen, S.~B., \& Fang, J.\ 2008, \apj, 676, 1210
\bibitem[Zirakashvili \& Aharonian(2007)]{Zirakashvili2007} Zirakashvili, V.~N., \& Aharonian, F.\ 2007, \aap, 465, 695


\end{thebibliography}
\end{document}